\documentclass[12pt]{article}
\usepackage[english]{babel}
\usepackage{amssymb,graphicx,wrapfig}
\author{Vladimir A. Petrov\footnote{e-mail: Vladimir.Petrov@ihep.ru}
}
\index{}
\title{On Nucleon "Radii"}
\date{}

\begin{document}

\maketitle
\begin{center}

A.A. Logunov Institute for High Energy Physics 

NRC "Kurchatov Institute", Protvino, RF
\end{center}
\begin{abstract}
We show that various hadron  "radii" that appear in the literature, e.g."charge", "gravitational", etc, do not have a direct geometric meaning and do not give an idea of the physical size of a hadron . They, however, can be used to estimate the genuine sizes of hadrons.

In the framework of the Gribov-Feynman  parton scheme we show that at high enough energy a gluon cloud appears beyond the valence core of the nucleon and begins to determine the region of interaction of elastic nucleon-nucleon scattering and the growth of its cross-section.
\end{abstract}

\section*{Introduction}

Since the pioneering experiments on the scattering of electrons by protons\cite{Ho} , which demonstrated first hand the non-point nature of protons, the concept of "charge radius" (along with its inevitable magnetic counterpart) has come into use in particle physics.
"Charge radius" appears in Particle Data Group publications as one of the most important physical characteristics of hadrons (both baryons and mesons), and, in the case of the proton, it has recently been the subject of lively discussion \cite{Me} regarding the correctness of its extraction from experimental data of various types.

It should be noted that quite often, "charge" is omitted from the name and it turns out, for example, simply "proton radius" (see e.g. Ref. \cite{Me}).

At first glance, since we are talking about the spatial distribution of electric charge carriers (quarks) inside the hadron, the "charge radius" fully characterizes the size of the region occupied on average by these carriers, and, therefore,
about the average size of the hadron as a whole.

However, as will be seen in the next section, things are not that simple.

\section{Is "charge radius" a radius?}
\subsection{The nucleon}

Without getting into the finer details \cite{Pe}  , let's take as an example the generally accepted definition of the (square of) "charge radius" of the nucleon:
\begin{equation}
r^{2}_{ch}= 6\frac{dF(t)}{dt}\mid_{t=0}
\end{equation}
with $ F (t) $ the electric form factor and $ t $, the transferred 4-momentum squared.
 If we take a PDG volume \cite{PDG}  , we find out that the "recommended" "charge radius" of the proton is
 \[r^{2}_{ch,proton}= (0.8414(19) fm)^{2}.\]
 
 For the neutron we find
 
\[r^{2}_{ch,neutron}= - 0.1155\pm 0.0017 fm^{2}.\]

Of course, it would hardly occur to anyone to interpret the last equality as meaning that the charge radius of the neutron is imaginary. Nevertheless, let's try to figure out why $ r^{2}_{ch,neutron} $ is negative.
 In work \cite{PO}  the following equalities were obtained
 
 \begin{equation}
 r^{2}_{ch,proton}= q_{u}\cdot N_{u}(proton)\cdot\langle r^{2}_{u}\rangle_{proton}  + q_{d}\cdot N_{d}(proton)\cdot \langle r^{2}_{d}\rangle_{proton};
 \end{equation}
 
 \begin{equation}
 r^{2}_{ch,neutron}= q_{u}\cdot N_{u}(neutron)\cdot \langle r^{2}_{u}\rangle_{neutron}+ q_{d}\cdot N_{d}(neutron)\cdot \langle r^{2}_{d}\rangle_{neutron}
 \end{equation}
where $ N_{a}(neutron) $ is the number of valence quarksv \footnote{Througout the paper we deal with current quarks.}of the type $ a $ in the nucleon, $ q_{a} $ stands for the quark $ a $ electric charge.
 The expression $ \langle r^{2}_{a}\rangle_{B} $ means the  average distance (squared) of the quark $ a $ in the nucleon $ B $ from its center.

 In the isotopic invariance approximation
 \[ \langle r^{2}_{u}\rangle_{neutron}=\langle r^{2}_{d}\rangle_{proton}\]
 \[\langle r^{2}_{u}\rangle_{proton} = \langle r^{2}_{d}\rangle_{neutron}. \]
 
 Thereof
 \begin{equation}
 r^{2}_{ch,proton}=\frac{4}{3}\langle r^{2}_{u}\rangle_{proton}-\frac{1}{3}\langle r^{2}_{d}\rangle_{proton}
 \end{equation}
 \begin{equation}
 r^{2}_{ch,neutron}= -\frac{2}{3}\langle r^{2}_{u}\rangle_{proton} + \frac{2}{3}\langle r^{2}_{d}\rangle_{proton}
 \end{equation}
 From equation (5) it is clear that negativity of $ r^{2}_{ch,neutron} $ is simply associated with the indeterminacy of the sign of electric charges, and itself is not a purely geometric quantity. On the contrary, the quantities $ \langle r^{2}_{u}\rangle_{proton} $ and $ \langle r^{2}_{d}\rangle_{proton} $ are the squares of the genuine radii averaged with positive probability distributions, which are, naturally, always positive.
 
 From Eqs.(4) and (5) one can estimate the physical (geometric)radii of the proton and neutron which appear equal in our approximation and read \cite{PO}
 
 \begin{equation}
 r^{2}_{proton} = r^{2}_{neutron} \equiv r^{2}_{nucleon}= r^{2}_{ch,proton}+ r^{2}_{ch,neutron}.
 \end{equation}
With the data on $r^{2}_{ch,proton}, r^{2}_{ch,neutron}$ recommended by the PDG we get
\begin{equation}
\sqrt{r^{2}_{nucleon}} \approx 0.77 fm.
\end{equation}

Let us compare this with
\[\sqrt{r^{2}_{ch,proton}}\approx 0.84 fm.\]

Thus, the "charge radius" is not a genuine radius, but is some kind of special construction that has no direct physical meaning, although it can be extracted from the fits of experimental data. In their turn, the physical radii can be extracted from the "charge" ones, as was shown above.
Last, but not least important, consideration. The "physical" radius of a nucleon, as is obvious from everything above, characterizes only the average size of the "habitat" of valence quarks. 
To especially emphasize this circumstance, we rewrite Eq. (7) as
\begin{equation}
r_{nucleon}^{val}= 0.77 fm.
\end{equation}

\subsection{The Kaon}

The form factor of the neutral pion is zero and the standard definition of the charge radius does not apply. 
 Let us consider instead the doublet $ K^{+}K^{0} $.
 
Form factors are:
\begin{equation}
F_{K^{+}}(t) = \frac{2}{3}\int dx u_{K^{+}}(x,t) + \frac{1}{3}\int dx \bar{s}_{K^{+}}(x,t)
\end{equation}
\[F_{K^{0}}(t) = \frac{1}{3}\int dx d_{K^{0}}(x,t) + \frac{1}{3}\int dx \bar{s}_{K^{0}}(x,t)\]
"Skew" parton densities $ u(x,t),... $ etc are defined as
\begin{equation}
f_{a}(x,t)= \int d^{2}b  \exp(i \textbf{q} \textbf{b}) \tilde{f}_{a}(x,\textbf{b}),
\end{equation}

$ \textbf{q}^{2}= -t $ and $ \tilde{f}_{a}(x,\textbf{b}) $ is the average number density of the type-$a$ parton in the nucleon ( longitudinal or light-cone)momentum fraction $ x $ and its transverse distance (from the nucleon center)  $ \textbf{b} $.

To continue in the same spirit as for nucleons, we (alas!) will make a very rough assumption about the exact $ SU(3)_{flavour} $ symmetry which gives
\begin{equation}
d_{K^{0}}=u_{K^{+}}, \bar{s}_{K^{0}}= \bar{s}_{K^{+}}.
\end{equation}
Under this condition, we obtain, using the values
\[r^{2}_{ch}(K^{+}) = (0.313\pm 0.001) fm^{2} \] and \[r^{2}_{ch}(K^{0}) = (-0.077\pm 0.010) fm^{2} \]
from \cite{PDG}, the following values for the physical sizes of kaons: 

\begin{equation}
r^{2}_{K^{+}} = r^{2}_{K^{0}} =0.275 fm^{2}=(0.525 fm)^{2}.
\end{equation}
 
Of course, taking into account the violation of unitary symmetry would lead to slightly different values of the radii.

Note also that despite the assumption of unitary symmetry in the expressions for the parton densities (10), the actual data from the PDG \cite{PDG} that we used, still introduce a difference in the average coordinates of $ u $ and $ \bar{s} $ quarks:

\begin{equation}
r_{u}^{2}(K^{+})= 0.39 fm^{2},   r_{\bar{s}}^{2}(K^{+})= 0.16 fm^{2}
\end{equation}
Heavier quarks lie "deeper". 

Let us also note that

\[r_{ch}(K^{+})/r_{ch}(proton)= 0.37.\]

At the same time we get for physical sizes

\[r_{K}/r_{nucleon} = 0.68 \approx 2/3.\]

which indicates that the approximate physical sizes of mesons and nucleons that we obtained are quite consistent with intuitive expectations.

\section{Other types of "radii" and what to do with them.}

As noted in the Introduction, other types of "radii" are also found in the literature. We will examine to what extent these characteristics can provide insight into the size of the gluon "cloud" of the nucleon. The need to introduce these new quantities is due to the fact that when working with electromagnetic form factors, we can only extract the radii associated with valence quarks, while gluons and sea quarks do not contribute and therefore cannot be taken into account.

 Let us take a certain entity that is sometimes called the "mechanical" or "gravitational"\cite{Ply} \footnote{The use of the term "gravitational" is not very relevant as it has long been used as a synonym for the Schwarzschild radius, which has a completely different physical meaning. Therefore, we will always use quotes.}  radius of the proton.
They are related not to the form factor of  electromagnetic current as in the case of the "charge radius" but to that of the density of the energy-momentum tensor.
Simplifying formalities(for greater clarity), we recall that both the "charge" and the "gravitational" ("mechanical") radii are obtained by differentiating the various moments of the "skew"("non-forward") parton distributions (10)
\begin{equation}
f^{a}_{J}(t)= \int dx x^{J-1}f_{a}(x,t)
\end{equation}

Note that $f_{a}(x,0)$ is a usual parton density,  $ f^{a}_{1}(0) = N_{a}$ is an average number of the type-$a$ constituents and $f^{a}_{2}(0) = \langle x_{a} \rangle $ is the average momentum fraction of the nucleon carried by \textit{ all}\footnote{ To obtain the average fraction per 1 parton, we do not have an experimentally accessible quantity related not to the \textit{density of the average number} of $a$-partons, $f_{a}(x,0)$, but to the \textit{probability density} of finding a parton $ a $ in the nucleon, $ w_{a}(x) $, normalized as $$ \sum_{a}\int dx w_{a}(x) = 1 $$ } the type-$ a $ constituents with a sum rule
$\sum_{a}\langle x_{a} \rangle =1 $ .

It is believed that the charge "radii" are obtained by differentiating linear combinations of the first   \;($J=1$) moments (with constituent charges as coefficients) (1), while for the the "gravitational" ones the form factors the second moments ($J=2$) are used.
So, the above mentioned "gravitational(mechanical)" form factor which we designate $ G(t) $\footnote{In fact the energy-momentum tensor containsa several form factors (4 for nuckeons , 2 for pions, etc). We deal here with that which is related to the momentum parton densities} is of the form
\begin{equation}
G(t) = \sum_{a = q_{v},gluons,q_{s}} f^{a}_{2}(t)
\end{equation}
where $ q_{v} $ and $ q_{s} $ mean valence quarks and "sea" $ q\bar{q} $ pairs.
By definition we get
\[G(0) = 1 = \sum_a f^{a}_{2}(0)= \sum_a  \langle x_{a} \rangle.\]

Let's now form, similar to Eq. (1), a "gravitational radius"

\begin{equation}
r^{2}_{G} =  6\frac{dG(t)}{dt}\mid_{t=0}.
\end{equation}

If this quantity corresponds to the usual idea of geometrical radius of a certain region?

Exactly as was done in the case of the electromagnetic form factor, we naturally come to partial "radii" $ r_{a}^{2}(x) $ defined as
\[6 \partial f^{a}(x,t)/\partial t \mid_{t=0}= r_{a}^{2}(x)f^{a}(x)\]

Thus we have
\begin{equation}
r^{2}_{G} = \langle\sum_{a} \sum_{i=1}^{n_{a}}  x_{i} r_{a}^{2}(x_{i})\rangle . 
\end{equation}
We see that "gravitational radius"$ r^{2}_{G} $ contains not only the information about
the spatial contribution of the nucleon constituents but also is "contaminated" by unnecessary information on their average fractions of the total momentum similar to the case of "charge radii" where spatial sizes were multiplies by the valence quark charge fractions. Note that the presence of the factor $ \langle x_{a} \rangle  $ in front of the squares of the partial radii in Eq.(12) apparently explains the smaller values of the "gravitational radii" compared to the "charge" ones.

Note that only in a particular degenerate case when the positions of constituents do not correlated with their momentum fraction and independent on the parton type $ a $ , i. e. when 
\begin{equation}
r_{a}^{2}(x) = r^{2}, \forall x , a ,
\end{equation}
we would have
(due to the sum rule $  \langle\sum_{a} \sum_{i=1}^{n_{a}}  x_{i} \rangle = 1 $)
\begin{equation}
r^{2}_{G} =  r^{2} .
\end{equation}
It is unlikely, however, that Eq.(13)and hence Eq.(14) hold.

As noted above, it is impossible to determine the average radii of the regions occupied by individual partons (valence and sea quarks and gluons) due to the lack of corresponding (skewed)probability densities $ w^{a}(x, t)$. The (skewed)parton  number density $ f^{a}(x,t) $ is not enough for this, as we see. 

The problem may be illustrated as follows. Lets us know the average value $ \langle\sum _{j=1}^{n} f(x_{j})\rangle $ and average $\langle n \rangle $ while we need the average $ \langle f(x)\rangle $ but the probability densities $ w_{n}(x_{1}, ...,x_{n}) $ are unknown to us.
The simplest thing that comes to mind is to try as an approximation for $ \langle f(x)\rangle $ the ratio
\begin{equation}
\bar{f} = \frac{\langle\sum _{j=1}^{n} f(x_{j})\rangle}{\langle n \rangle}.
\end{equation}

Again, if the number $ n $ does not fluctuate , $ \langle n \rangle = n =fix $ then
\[\bar{f} = \langle f \rangle.\]
Obviously, this does not apply to gluons and sea pairs $ \bar{q}q $, whereas it does apply to valence quarks.

Let us recall that both quantities F and G, being defined by the matrix elements of conserved operators, $ J_{\mu} $ and $ \Theta _{\mu\nu} $, do not depend on the scale of renormalization. This, however, is not true for quantities like $ \langle x_{a} \rangle  $. 

Thus, unlike the case with the electromagnetic form factor, where we were able to estimate the physical size of the nucleon's valence core using data on the proton and neutron form factors (in the approximation of exact isotopic symmetry), the form factors associated with the energy-momentum operator do not seem to give us a simple way to estimate the average physical size of the gluon and sea quark habitat in the nucleon.

\section*{Form Factors and  Sizes.}

Above we tacitly assumed that the definition of "radii" through the derivative of the form factor (see Eqs. (1), (11)) borrowed from non-relativistic quantum mechanics remains valid in the general case as well. It is interesting to see what the "coordinate content" of these definitions is in terms of correlations of quantum field operators.

For definiteness, let us take the form factor of the baryon current operator
\[J_{\mu}^{B} = \sum_a \bar{\psi}_{a}\gamma_{\mu} \psi_{a}.\]

Baryon  number of the nucleon is given by the corresponding form factor $ B(t) $ and evidently 
\[B(0) = 1.\]

If to define the "baryon radius" $ r_{B} $ via usual expression
\begin{equation}
r_{B}^{2} = 6\frac{dB(t)}{dt}\mid_{t=0}
\end{equation}
then it appears to coincide with the \textit{physical} "valence quark" radius of the nucleon in Eq.(6)  because, in contradistinction to the electromagnetic form factor, it does not contain quark electric charges
\[r_{B}^{2} = \frac{2}{3}r_{u}^{2} + \frac{1}{3}r_{d}^{2} =r^{2}_{nucleon}= r^{2}_{ch,proton}+ r^{2}_{ch,neutron}.\]
here factors $ 1/3 $ are related to the quark baryon numbers while the factors $ 2/3 $ and $ 1/3 $ acquire the meaning of the probabilities to find $ u $ and $ d $ quarks in the proton.

Experimental measurement of the baryonic form factor would be possible if one could find a process where the operator $ J_{\mu}^{B} $ could play the role of composite operator in a Wilson product expansion. This would be the most relevant case as it would give us not the "charge" but the physical size of the nucleon. We, however, could not find such a process.

Anyway, we have reproduced Eq.(6)( as it should be) but in a more straightforward way.

To reveal the coordinate content of Eq.(16) we apply the Bogolyubov \cite{Bog} reduction techniques according to which
we get\footnote{ $\vert 0\rangle $ and $\vert p\rangle $ designate the vacuum and one-nucleon states, correspondingly; $ m $ stands for the nucleon mass and we mean averaging in nucleon polarizations.} 
\begin{center}
$ B(t = {q}^{2}) = \frac{2p_{\mu}}{4m^{2}- {q}^{2}}\int d^{4}x e^{iqx} \langle 0\vert \frac{\delta J^{B}_{\mu}(x)}{\delta \bar{N}(0)}\vert p \rangle $  
\end{center}
Variation derivative is taken over the nucleon  out-field $ N(x) $.
\begin{center}
$ \frac{\delta J_{\mu}^{B}(x)}{\delta\bar{N}(0)}= i\theta (-x)[J^{B}_{\mu}(x),\eta_{N}(0)]. $
\end{center}
Here $\eta_{N}(x) = i\frac{\delta S}{\delta \bar{N}(x)}S^{+} $ is the nucleon density operator.
For definiteness let us take the laboratory frame where $ \textbf{p} =0 $.
We get 
\begin{center}
$ B(q^{2}) = \frac{2m}{4m^{2}- q^{2}}\int d^{4}x \exp [ -i\frac{q^{2}x^{0}}{2m} -i(\textbf{xn})\sqrt{-q^{2}(1-q^{2}/4m^{2})}] \langle 0\vert \frac{\delta J^{B}_{0}(x)}{\delta \bar{N}(0)}\vert \textbf{p}=0 \rangle$.
\end{center}
Now, if we apply to this representation the formula (20) for the "baryon radius"
we have
\[\textbf{r}^{2}_{B} = \int d\textbf{r}\textbf{r}^{2}\rho_{B}(\textbf{r})\]
where
\begin{equation}
\rho_{B}(\textbf{r}) = \frac{1}{2m}\int dx^{0}\langle 0\vert \frac{\delta J^{B}_{0}(x^{0},\textbf{r})}{\delta \bar{N}(0,\textbf{0})}\vert \textbf{p}=0 \rangle
\end{equation}
is to have the meaning of the net baryon density inside the nucleon at rest.

For the "charge radius" we get similar expressions
\begin{equation}
\rho_{ch}(\textbf{r}) = \frac{1}{2m}\int dx^{0}\langle 0\vert \frac{\delta J_{0}(x^{0},\textbf{r})}{\delta \bar{N}(0,\textbf{0})}\vert \textbf{p}=0 \rangle
\end{equation}
and
\begin{equation}
\textbf{r}^{2}_{ch} = \int d\textbf{r}\textbf{r}^{2}\rho_{ch}(\textbf{r})
\end{equation}

Quantities $ \textbf{r}^{2}_{B} $ and $ \textbf{r}^{2}_{ch} $ are Lorentz invariant (according to their definition as derivatives of form factors) but their interpretation as average "radii" refers to the rest ("lab") frame of the nucleon.

Note that nothing prevents us from giving a more general definition of the charge density in a moving nucleon,viz.,
\begin{equation}
\rho_{ch}(\textbf{r}|\textbf{p}) = \frac{1}{2\sqrt{m^{2}+p^{2}}}\int dx^{0}\langle 0\vert \frac{\delta J_{0}(x^{0},\textbf{r})}{\delta \bar{N}(0,\textbf{0})}\vert \textbf{p}\rangle
\end{equation}
Note that
\[\int d\textbf{r}\rho_{ch}(\textbf{r}|\textbf{p})= Q, \forall \textbf{p}, \] (nucleon charge).
We,  however, will not develop this subject here.

When looking at Eq.(18) we notice that the distance $ |\textbf{r}| $ relates points taken at different times $ x^{0}, 0 $ and so the profile of the supposed charge distribution $ \rho_{ch}(\textbf{r}) $ doesn't give us an instantaneous snapshot of the charge distribution inside the pion but rather \textit{something smeared in time}.

One can prove (with use, e.g., of the Jost-Lehmann-Dyson representation for causal commutators, see e.g. \cite{Bog} ) that in the non-relativistic limit ( $ c $ is the speed of light, $ c\rightarrow\infty $)
\begin{center}
$ \langle 0\vert \frac{\delta J_{0}(ct,\textbf{r} )}{\delta \bar{N}(0,\textbf{0})}\vert \textbf{p}=0 \rangle |_{c \longrightarrow \infty} = \delta (t) \Phi (\textbf{r}) .$
\end{center}
So we recover a NR quantum-mechanical equal-time case and the above expression  gives an instant snapshot of the object.

On the other hand, one can argue that non-simultaneity in the definition of the baryon radius ( or a nucleon size) can be taken into account if to assume that arising uncertainty is given by the "retardation time"  $ \sim \left\langle r\right\rangle/c $ and may induce an uncertainty comparable with the very radius in question \cite{Pe}.

On general grounds we were no able to prove (or disprove) the positivity of the function $ \rho_{B}^{lab}(\textbf{r}) $, a  necessary property of a number density.

It should be noted that the clear mathematical meaning of the spatial parameter $ \textbf{r} $ that naturally arises in our approach differs from the meaning of the "distance" parameter which was \textit{introduced }, e.g. in Refs.\cite{Me} and \cite{Ply} "by hand".
The authors of \cite{Me} consider matrix elements of the current or energy-momentum tensor in the Breit frame 
\[p' = (\sqrt{m^{2}+\textbf{q}^{2}/4}, \textbf{q}/2),\;p =(\sqrt{m^{2}+\textbf{q}^{2}/4}, -\textbf{q}/2) \]
and then integrate it with 
\[\int d\textbf{q}  exp(i\textbf{q}\textbf{r})/(2\sqrt{m^{2}+ \textbf{q}^{2}/4} (2\pi)^{3})\]
thereby introducing an arbitrary spatial parameter $ \textbf{r} $ without any connection with the coordinate dependence of the 
relevant field operators. 

The difference is evident if we compare, e.g. the expressions for $ \rho_{ch} (\textbf{r}) $.
In fact, our expression (17) can be represented in a more symmetrical way:
\begin{equation}
\rho_{ch} (\textbf{r}) =\int d^{4}Xd^{4}\xi R_{0}(X,\xi)[\delta (\textbf{X}-\textbf{r}) \exp im \xi_{0}];
\end{equation}
while the definition used in \cite{Me} leads to (we use a modified designation $ \tilde{\rho} $ for this case)
\begin{equation}
\tilde{\rho}_{ch} (\textbf{r}) =\int d^{4}Xd^{4}\xi R_{0}(X,\xi)[2iD^{-}(\xi_{0}/2,\textbf{X}-\textbf{r})]
\end{equation}
where
\[R_{0}(X,\xi)= \frac{1}{2m}\langle 0\vert \delta^{2}J_{0}(X)/\delta \bar{N}(\xi/2)\delta N (-\xi/2)\vert 0\rangle\]
and
\[D^{-} (x_{0},\textbf{x})= \frac{i}{(2\pi)^{3}}\int d^{4}k e^{ikx} \theta (-k_{0})\delta (k^{2}-m^{2}).\]
Densities $ \rho_{ch} (\textbf{r}) $ and $ \tilde{\rho}_{ch} (\textbf{r}) $ lead to the same $ r_{ch}^{2} $  but it does not mean that they are equal to each other,the higher moments of $\textbf{ r} $ will be different.
Note that the expression for $ \rho_{ch} (\textbf{r}) $ is \emph{derived} from the general rules of q.f.t. reduction, while the expression for $ \tilde{\rho} $  is essentially\emph{ postulated}.

Strictly speaking, based on the spatio-temporal considerations of the theory of relativity \cite{Lg} , the measure of size should contain
\[R_{0}(X=(0,\textbf{X}),\xi).\]
However, at the moment the author does not quite understand how to do this without a priori postulation.

Despite the fact that we have managed to somewhat clarify the field-theoretical meaning of the "radii" discussed above (see Eq.(17)), the only result that remains the same is the estimate of the physical radius of the "valence core" of the nucleon, based on the form factors of the proton and neutron. This, as noted above, does not allow for gluons and sea quarks to be taken into account. Using the energy-momentum tensor operator instead of the electromagnetic (or baryon) currents also does not allow, as we have seen, for a direct estimate of the dimensions associated with the gluon field.However, approximately, considering the contribution of valence quarks to be a fortiori suppressed, this is possible, as will be seen below.

For lack of a better way, we will now try to fill this gap, at least a little, using a simplified phenomenological model for "skewed" (non forward) parton distributions.

\section*{When Do Gluons Come Out?}

Above we mainly concerned the valence quarks. Certainly, hadrons consist not only of valence quarks, but also of gluons and "sea" quark-antiquark pairs. Especially since it is argued that, say, the nucleon mass is mainly formed by the gluon field.
Thus, the question about the size of the habitat area of "intranucleon gluons" is also not idle.
In the previous Section the "radii" were referred to the rest frame of the nucleon. In particular, the transverse size \textit{of the valence core} of the nucleon remains frame independent.

Actually quantum fluctuations and their life times and spatial extents are not Lorentz invariant\footnote{This in no way affects the Lorentz invariance of the observed cross sections.} . For example, the "life time" $ \Delta t  $ of a fluctuation of the nucleon with  high momentum $ P $ into a parton system with $ n $ partons is (with evident designations)

\[\Delta t = 2P/\sum_{i=1}^{n} (m_{\perp, i}^{2}/x_{i} - m^{2}).\]
It is general consent that high-energy nucleon-nucleon  scattering at higher energies is defined by the overlap of their "gluon clouds"\footnote{This also applies to "sea" $ q\bar{q} $ pairs, but in the following we will focus only on gluons, assuming that the $ q\bar{q} $ pairs generated by them are included in the effective distributions.}  which swell with growth of the collision energy.
So, it is natural to compare the (transverse projection) of the size of the  valence core with the spatial extent of the gluon content of the \textit{fast moving nucleon}. It is believed that in the nucleon rest frame the gluon field is "hidden" in the valence core.

The following considerations are closely related to the gluon number density on the basis of the following phenomenological model for its  "skew" generalization:
\begin{equation}
g(x,t) = c(1/x)^{\alpha_{\mathcal{P}} (t)} (1-x)_{+}^{5}.
\end{equation}
which stems from the definition of $ g_{J}(t) $, the $ Jth $ Mellin moment of the corresponding twist-2 gluon composite operator
\begin{equation}
 \langle p+q\vert O_{g}^{\mu_{1},...,\mu_{J}}\vert p\rangle = \Pi ^{\mu_{1},...,\mu_{J}}(q,p)g_{J}(t), t = - q^{2}
 \end{equation} 
and assuming the dominance of the Pomeron pole at $ J= \alpha_{\mathcal{P}}(t)\approx 1 + \Delta + \alpha'_{\mathcal{P}} t +... $ in the $ J $-complex plane. The factor $ (1-x)_{+}^{5} $ results from the constituent counting rules \cite{Mur}  according to which the parton density $ \sim (1-x)_{+}^{2n_{spectators} -1} $ at $ x\rightarrow 1 $. In the case of the gluon number density $n_{spectators} $ is evidently equal to $ 3 $, the total number of the valence quarks in the nucleon. Further on we  will not care about the RG evolution considering the RG scale "frozen" because when constructing physical, measurable amplitudes, the dependence on the RG scale should disappear anyway. In the case of the valence core, it is no longer present already at the level of partonic distributions determined by conserved currents of the type $ J_{\mu}^{el.mag.} $ or $ J_{\mu}^{B} $.

According to Gribov's arguments \cite{Gr}, the lower limit of integration over $ x $ is $ \Lambda/P $ where $ P $ is the nucleon momentum
while $ \Lambda $ is some minimum value of the parton longitudinal ( or light-cone) momentum. 

Let us now consider the joint parton distributions both in longitudinal momentum $ x $ and transverse (w.r.t. to the direction of the nucleon momentum) 2D coordinate $ \textbf{b} $ (impact parameter)
\begin{equation}
\tilde{g}(x,\textbf{b}) = cg(x,0)\exp(-b^{2}/R^{2}(x))/\pi R^{2}(x).
\end{equation}
where \[R^{2}(x\ll 1)\approx 4\alpha'_{\mathcal{P}}ln(1/x) + b_{0}^{2}  .\]
This is just Fourier-Bessel transform of Eq.(18).

From this requisite we obtain (after averaging in $ x $)\footnote{It should be reminded again that in the absence of a parton \textit{probability density} (not to be confused with the \textit{average parton number density}) we are forced - hoping that the fluctuations are not too large - to use a rough estimate of the type presented in Eq.(20).}  that the gluon field on average occupies in the transverse plane (of a nucleon flying with a large momentum $ P\gg\Lambda $) an area
\begin{equation}
4\pi \alpha'_{\mathcal{P}}\gamma (\Delta ln(P/\Lambda))/\Delta
\end{equation}
where 
\[\gamma (z)= (z e^{z}-c')/(e^{z}-c) - 1 \]
with
\[c = 6!\Gamma(1-\Delta)/\Gamma(6-\Delta), c' = c \Delta( \psi(6-\Delta) - \psi( 1-\Delta)), \psi(z)= \Gamma^{'}(z)/\Gamma(z).\]
The function $ \gamma (z) $ grows monotonically as $\approx z $ at $z \rightarrow \infty. $

Let us compare the "gluon habitat" with the (transverse) area occupied by the "valence core", i.e.
\begin{center}
$ \pi b_{N}^{2} \equiv \frac{2}{3}\pi (r_{nucleon}^{val})^{2} .$
\end{center}
The gluon field begins to "crawl out" from the valence core when the nucleon momentum $ P $ reaches the " critical" value $ P^{*}
 $ which can be found from a transcendental equation
 
\[\gamma (\Delta ln(P^{*}/\Lambda)) = \Delta (b_{N}^{2} - b_{0}^{2})/4\alpha_{\mathcal{P}}^{'}.\]

For a qualitative  estimate, let us take the parameters from our paper  \cite{PO}:
\begin{center}
$ \alpha'_{\mathcal{P}} = 0.234\; GeV^{-2},b_{0} = 2.84\;
  GeV^{-1} , \Delta =0.102 .$
\end{center} 
As to the parameter $ \Lambda $, we are to reconcile it with the energy scale $ \sqrt{s_{0}} $ at which the above parameters were obtained in Ref.\cite{PO} where $ s_{0} $ was taken $1\; GeV^{2}  $, i.e. we have 
\[\sqrt{s/s_{0}}= 2P/\sqrt{s_{0}}\]
and therefore 
\begin{center}
$\Lambda \approx 0.5 \;GeV.  $
\end{center}
Then we get
\[P^{\ast} \approx  12 \; GeV.\]
This would correspond to a collision of two protons at a center of mass energy 
\begin{center}
$ \sqrt{s} \approx  24 $ GeV\footnote{Due to the exponential dependence on the parameters, this estimate should not be taken too literally due to its significant model dependence. However, the qualitative prediction of gluon "crawling out" certainly remains valid.} .
\end{center}

Thus, from our estimate it follows that the growth of the \textit{elastic} cross section of $ pp $ scattering is obviously associated with the beginning of the gluon field creeping beyond the valence core.

For quantitative arguments we can take the COMPETE parametrization \cite{COMP} which  shows that the $  pp $ \textit{total}   cross-section begins to grow in the vicinity of $ \sqrt{s}  = 10.69 $ GeV. It also follows from the fit used by TOTEM  \cite{TOT}  that the \textit{elastic} $  pp $ cross-section reaches its minimum at $ \sqrt{s} =19.58 $ GeV, while according to ATLAS \cite{AL}  this occurs at $ \sqrt{s} = 22.76 $ GeV.
We find that this confirms our arguments given above and the growth of elastic cross sections begins when the valence cores of colliding nucleons cease to overlap, giving way to swelling gluon "coats". 

What does explain the earlier start of growth of total cross sections compared to elastic ones? It is evidently the fact that the ever rising (due to the exponentially increasing number of opening channels) inelastic cross section masks the decrease of the elastic one: the derivative of the total cross-section vanishes due to mutual cancellation of the positive rate of the inelastic and still the negative rate of the elastic cross sections.  And only after passing the minimum in the region of about $ 20 $ GeV the $ pp $ elastic cross section begins to make a positive contribution to the growth of the $ pp $ total cross sections.

\section*{Conclusions}

We have considered  the notion of the nucleon size and tried to clarify its field theoretic content.
Physical (geometrical) values of the nucleon and kaon sizes are derived.
In the framework of the Gribov parton scheme we have shown that
at high enough energy the gluon cloud of the nucleon shows up and going beyond the valence core begins to define the nucleon-nucleon interaction region.
It turns out that it occurs near the energies when the \textit{elastic} $ pp $ cross-section begins to rise.
We believe that in terms of the interplay between the sizes of the valence core of a nucleon and its gluon cloud discussed above, this effect, which clearly relates to the peripheral interaction, receives an interesting and visual physical content and interpretation.
Regarding the specific and quite a popular direction in the study of the internal properties of hadrons, initiated in the works of M. Polyakov (excellent review of his contributions is given, for example, in the work of C. Lorc\'{e} and P. Schweitzer \cite{Me} ), the arguments presented in the given paper, from a formal point of view, do not contradict the research in terms of pressures, stresses, shear viscosities,etc  which are working concepts in the analysis of the properties of continuous media (solid and liquid) in the theory of resistance of materials. However, I do not see the adequacy of these concepts for discrete quantum structures.

 \section*{Acknowledgements}
 
 I am grateful to N.P. Tkachenko, A. K. Likhoded, V.T. Kim and S. M. Troshin for interest to my talk and fruitful discussions.

\end{document}